\documentclass[a4paper]{article}
\usepackage{ISCSLP2026}
\usepackage{ifthen}

\usepackage{enumitem}
\usepackage{comment,multirow}
\usepackage[table]{xcolor}
\newboolean{blind}
\setboolean{blind}{false} 
\title{Is Semantics Enough for Speech Mean Opinion Score Prediction?}
\name{
Tianyu Lan,
Yufei Shi,
Yang Ai$^{*}$,
Honghao Sun,
Huipeng Du,
Zhenhua Ling
\thanks{$^{*}$Corresponding author. This work was funded by the National
Natural Science Foundation of China under Grants U23B2053 and 62301521.}
}
\address{
National Engineering Research Center of Speech and Language Information Processing,\\
University of Science and Technology of China, Hefei, China
}

\email{
lantianyu@mail.ustc.edu.cn,
zkddsr2023@mail.ustc.edu.cn,
yangai@ustc.edu.cn,\\
honghaosun@mail.ustc.edu.cn,
redmist@mail.ustc.edu.cn,
zhling@ustc.edu.cn
}

\begin{document}

\maketitle
\begin{abstract}
 Mean Opinion Score (MOS) is the gold standard for evaluating synthesized speech naturalness. However, current automatic MOS predictors are dominated by self-supervised learning (SSL) models that prioritize high-level semantics, potentially compromising their ability to capture critical acoustic details. In this paper, we systematically investigate representations from three paradigms: SSLs, acoustic-only neural audio codecs (NACs), and unified NACs that integrate semantics into reconstruction-based architectures. Extensive benchmarking on the standard BVCC and multiple out-of-domain (OOD) datasets demonstrates that features synergizing semantic understanding with fine-grained acoustic modeling achieve a higher performance upper bound in speech quality assessment. Ultimately, our findings highlight that semantics alone are not enough; a dual focus on semantic content and acoustic fidelity is essential for robust MOS prediction.
\end{abstract}


\section{Introduction}

The rapid evolution of generative speech technologies, including text-to-speech (TTS) and voice conversion (VC), has created a critical need for efficient evaluation methods. While the human-rated mean opinion score (MOS) remains the gold standard for assessing the naturalness of speech \cite{black2005blizzard}, manual annotation is prohibitively costly. Feature extraction is a crucial component of MOS prediction models. Current speech representations can be broadly categorized into two main domains: \textbf{Semantic Representations} extracted by self-supervised learning (SSL) models \cite{mosanet,saeki2022utmos,sslmos,tseng2022ddos,kunevsova2023ensemble} (such as \textbf{Wav2Vec 2.0} \cite{baevski2020wav2vec}, \textbf{HuBERT} \cite{hsu2021hubert}, and \textbf{WavLM} \cite{chen2022wavlm}), and \textbf{Acoustic Representations} derived from low-level representations \cite{patton2016automos,fu2018quality,lo2019mosnet} (such as raw waveform and mel-spectrogram) and Neural Audio Codecs (NACs) (such as \textbf{EnCodec} \cite{defossez2022encodec} and \textbf{DAC} \cite{kumar2023DAC}). Most current state-of-the-art MOS prediction models rely predominantly on semantic-based SSL models as their feature extraction backbones \cite{saeki2022utmos,wang2023ramp,vioni2023investigating,baba2024utmosv2}.

These models have demonstrated success, largely because human quality ratings are strongly correlated with high-level semantics, a feature that SSL models capture exceptionally well through contextual representation learning. However, this reliance on SSL presents a limitation. The pre-training objectives of SSL models (e.g., masked prediction) are designed to abstract high-level contextual semantics, a process that inherently encourages the model to overlook fine-grained acoustic details, such as environmental noise and waveform distortions. This divergence raises a fundamental research question: \textbf{Is Semantics Enough for Speech Mean Opinion Score Prediction?} While semantics are dominant, we hypothesize that the neglect of acoustic details in SSL creates a ``blind spot'' for perceptually significant artifacts, thereby imposing a ceiling on the representation's upper bound for quality assessment. This limitation is particularly pronounced in naturalness MOS prediction, where current SSL-based SOTA models still leave significant room for improvement in utterance-level metrics.

To rigorously investigate this hypothesis, we move beyond standard fine-tuning and evaluate predictors in frozen-encoder settings to probe the intrinsic information content of representations. In this study, \textbf{we conduct the first comprehensive comparative study of representations derived from three categories}: (1) \textit{SSLs} (Wav2Vec 2.0, HuBERT and WavLM) as strong baselines; (2) \textit{acoustic-only NACs} (EnCodec and DAC); and (3) \textit{unified NACs} (e.g., XCodec \cite{ye2025xcodec}, SpeechTokenizer \cite{zhang2023speechtokenizer}) that integrate semantics with acoustic reconstruction. We conduct our primary training and in-domain evaluation on the BVCC \cite{cooper2021BVCC} dataset. Furthermore, to rigorously assess robustness against diverse distribution shifts, we extend the evaluation to two distinct OOD datasets: SOMOS \cite{maniati2022somos} (cross-corpus, matched-language) and BC2019 \cite{wu2019BC2019} (cross-corpus, cross-lingual). Covering a broad spectrum of distribution shifts, this suite serves as a comprehensive testbed to analyze the generalization capabilities of different representations.

\begin{figure*}[t]  
    \centering
    \includegraphics[width=0.90\textwidth]{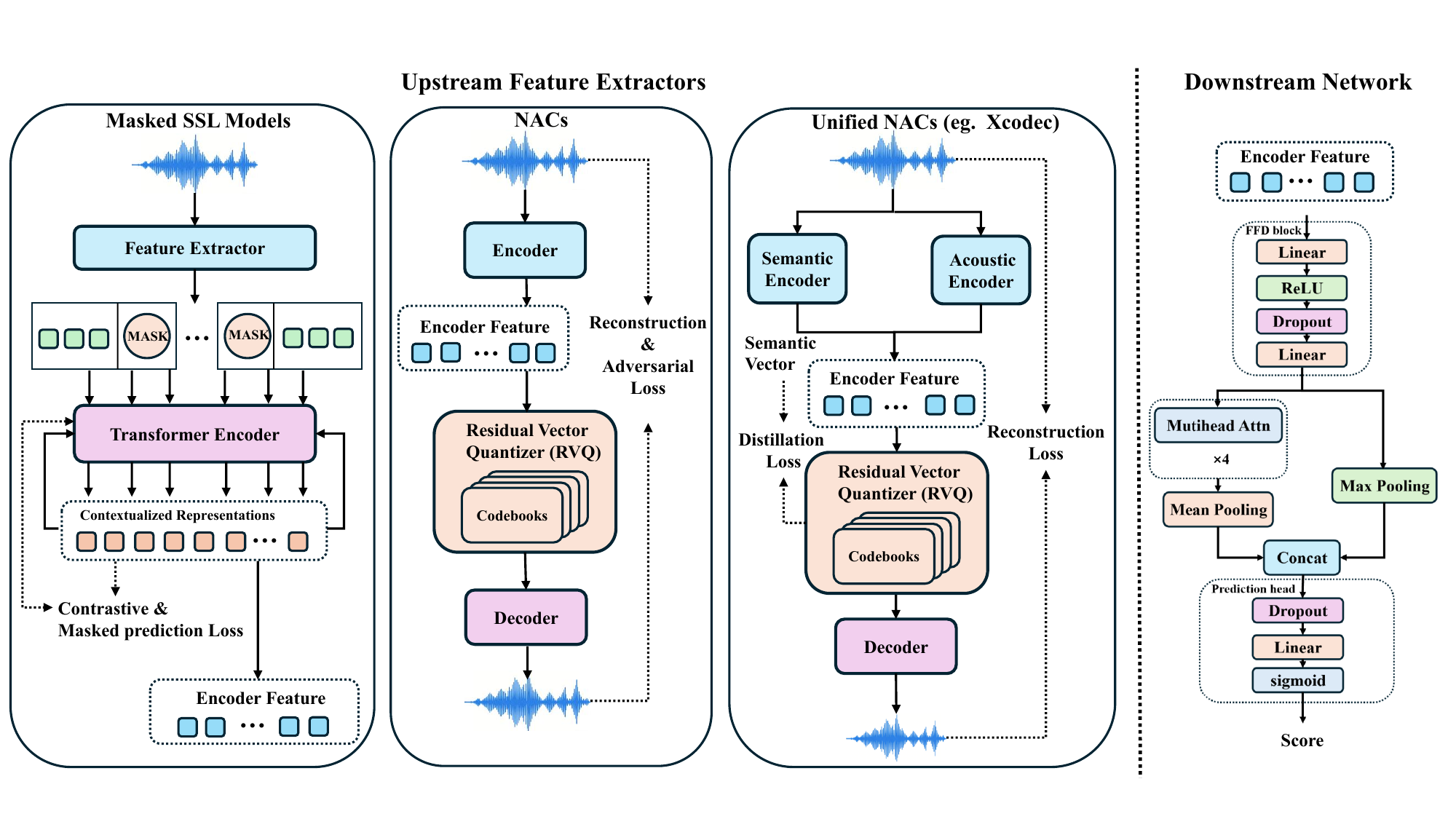} 
    \caption{The architecture of the Upstream Feature Extractors and Downstream Prediction Network.}
    \label{fig1}
\end{figure*}

Our experimental results provide a compelling answer to the central question: \textbf{semantic representations alone are not enough for high-fidelity speech MOS prediction.}  We demonstrate that representations synergizing semantic understanding with fine-grained acoustic modeling consistently establish a higher performance upper bound than standard semantic-focused SSLs or acoustic-only baselines. In matched-language settings, this advantage persists across both frozen and fine-tuning settings, validating the value of acoustic priors as a critical complement to semantics. While semantic optimization during fine-tuning tends to narrow the performance gap, the superior initialization of unified models remains advantageous. A notable exception, however, arises in cross-lingual OOD scenarios: for unseen languages, predictors utilizing SSL features exhibit comparable or even superior generalization compared to unified NACs.

\section{MOS Prediction Framework}

To investigate the role of different representations in MOS prediction, as illustrated in Figure 1, we adopt a two-stage MOS prediction architecture comprising an upstream feature extractor and a downstream prediction network. We posit that the pre-training objective of the feature extractor primarily determines the information content of its latent representations. In this section, we first classify the feature extractors into three paradigms based on their objectives. Subsequently, we describe the unified downstream prediction network used across all experiments and define our investigation protocols.

\subsection{Upstream Feature Extractors}

We select models from three distinct paradigms to serve as the upstream feature extractors. To evaluate the information content of these representations without information loss from quantization, we extract the continuous, unquantized latent vectors from the encoders—namely, the hidden states for SSLs and the pre-quantization vectors for codecs. Detailed configurations of these models are summarized in Table \ref{table:model_config}.

\begin{table}[t]
  \centering
  \scriptsize 
  \renewcommand{\arraystretch}{0.65} 
  \setlength{\tabcolsep}{4pt} 
  
  \caption{Model configurations and parameters. "Ext. Layer" is the feature extraction layer; "Pre-train" indicates the total duration of pre-training data.}
  \label{table:model_config}
  \vspace{-2mm} 
  
  \begin{tabular}{lcccc} 
    \toprule
    \textbf{Model} & \textbf{Frame Rate} & \textbf{Params} & \textbf{Ext. Layer} & \textbf{Pre-train} \\
    \midrule
    w2v2\_base      & 50 Hz & 95M  & 12   & 960h   \\
    hubert\_base    & 50 Hz & 95M  & 12   & 960h   \\
    hubert\_large   & 50 Hz & 316M & 24   & 60000h \\ 
    wavlm\_base     & 50 Hz & 95M  & Avg. & 84000h \\
    \midrule
    EnCodec         & 75 Hz & 7M   & 1    & 16000h \\
    DAC             & 50 Hz & 22M  & 1    & 20000h \\
    \midrule
    Xcodec-hubert   & 50 Hz & 123M & 1    & 960h   \\
    Xcodec-wavlm    & 50 Hz & 123M & 1    & 46000h \\
    SpeechTokenizer & 50 Hz & 68M  & 1    & 960h   \\
    \bottomrule
  \end{tabular}
  \vspace{-4mm} 
\end{table}

\noindent\textbf{SSL Models.}
We select Wav2Vec 2.0 \cite{baevski2020wav2vec}, HuBERT \cite{hsu2021hubert}, and WavLM \cite{chen2022wavlm} as representative baselines. These models are typically trained with masked prediction objectives. The core mechanism forces the model to infer masked temporal segments based on the surrounding context. To succeed in this task, the model must abstract high-level semantic structures and global dependencies. Empirically, comprehensive benchmarks like SUPERB \cite{yang2021superb} validate this behavior: SSLs excel in semantic-heavy tasks (e.g., speech recognition) but are heavily sub-optimal for raw acoustic reconstruction. Layer-wise analyses of models like HuBERT further confirm that these objectives inherently encourage the abstraction of phonetics while deliberately discarding specific, fine-grained acoustic details \cite{pasad2021layer}. Consequently, this paradigm results in a strictly semantic-dominant representation.

\noindent\textbf{Acoustic-only NACs.} We select EnCodec \cite{defossez2022encodec} and DAC \cite{kumar2023DAC}. These models employ an encoder-decoder architecture trained end-to-end. The training is driven by a combination of reconstruction loss and adversarial loss. Unlike SSLs, the encoder here is strictly penalized if the reconstructed audio deviates from the original waveform. This forces the latent representation to preserve fine-grained acoustic details, including phase and pitch. However, without an explicit semantic modeling objective, these representations lack high-level semantic coherence, resulting in a primarily acoustic-dominant representation. This clear dichotomy between semantic SSLs and acoustic NACs has been explicitly validated and leveraged in generative systems like AudioLM \cite{borsos2023audiolm}, which depend on NACs specifically for acoustic detail preservation.

\noindent\textbf{Unified NACs.} We select SpeechTokenizer \cite{zhang2023speechtokenizer} and Xcodec \cite{ye2025xcodec}. These models integrate the reconstruction objective of traditional codecs with a semantic modeling objective. Specifically, they leverage representations from pre-trained SSL models to guide the codec's encoding process, employing mechanisms such as knowledge distillation or explicit feature fusion. By imposing these compound constraints, unified codecs attempt to inject semantic understanding into the acoustically rich latent space (or align a subspace with semantic priors). Our analysis focuses on whether this explicit integration successfully creates a comprehensive unified representation suitable for MOS prediction.

\subsection{Downstream Prediction Network}
To ensure that performance differences are primarily attributable to the input representations, we employ a consistent prediction network across all experiments. Given the input feature $\mathbf{H} \in \mathbb{R}^{T \times D}$ derived from the upstream feature extractor—where $T$ and $D$ denote the temporal and feature dimensions, respectively—the prediction network first applies a feed-forward network (FFN) to obtain an intermediate representation $\mathbf{H}_{\text{FFN}} \in \mathbb{R}^{T \times D}$. This is followed by a stack of four multi-head self-attention layers to capture context-aware dependencies, yielding $\mathbf{H}_{\text{attn}} \in \mathbb{R}^{T \times D}$. We adopt a hybrid pooling strategy that concatenates the global mean-pooled representation of $\mathbf{H}_{\text{attn}}$ with the global max-pooled representation of $\mathbf{H}_{\text{FFN}}$:
\begin{equation}
\mathbf{h}_{\text{pool}} = \text{Concat}(\text{MeanPool}(\mathbf{H}_{\text{attn}}), \text{MaxPool}(\mathbf{H}_{\text{FFN}})).
\end{equation}
Finally, the aggregated vector $\mathbf{h}_{\text{pool}}$ is passed through a linear regression head with a Sigmoid activation, which is explicitly scaled to the MOS range of $[1, 5]$:
\begin{equation}
\hat{y} = 1 + 4 \cdot \sigma(\mathbf{W}\mathbf{h}_{\text{pool}} + b).
\end{equation}
where $\mathbf{W}$ and $b$ denote the learnable weight matrix and bias of the linear head, and $\sigma$ is the Sigmoid function.

\subsection{Investigation Protocols}

We design two experimental settings to evaluate the predictors:

\begin{itemize}[leftmargin=*,labelsep=5pt]
\item \textbf{Frozen Setting.} The upstream feature extractor is frozen, and only the downstream network is trained. This acts as a diagnostic probe to evaluate the intrinsic information content of the pre-trained representations, ensuring a fair comparison of raw capabilities without task-specific adaptation.

\item \textbf{Fine-Tuning Setting:} Upstream encoders (both SSL and NAC) are fully unfrozen for end-to-end optimization. This allows gradients from the MOS regression objective to propagate through the entire network.
\end{itemize}

\section{Experimental Setup and Results}
\subsection{Datasets}
We utilize the BVCC \cite{cooper2021BVCC} dataset for primary training and in-domain evaluation, and SOMOS \cite{maniati2022somos} and BC2019 \cite{wu2019BC2019} for zero-shot evaluation. Across all datasets, subjective ratings range from 1 to 5, and the mean rating of each sample is used as the ground-truth MOS. Details of the three datasets are as follows

\noindent\textbf{BVCC.} The BVCC dataset contains 7,106 English utterances, with the training/validation/test sets split in a ratio of 70\%/15\%/15\%. The data comes from participating systems from past Blizzard Challenges (BCs), the Voice Conversion Challenges systems, and samples generated by ESPNet \cite{watanabe2018espnet}. Each sentence in BVCC has scores from 8 listeners.

\noindent\textbf{BC2019.} The BC2019 dataset contains Mandarin Chinese TTS samples from the Blizzard Challenge 2019. We utilized the official test set comprising 540 utterances for zero-shot evaluation. Each sample includes 10 to 17 ratings from listeners.

\noindent\textbf{SOMOS.} The SOMOS dataset consists of English neural TTS samples generated by 200 systems with prosodic variations. We utilized the 3000 samples from SOMOS-clean test subset. Each utterance is evaluated by at least 17 listeners.

\subsection{Implementation Details and Evaluation metrics}
All models are implemented using PyTorch. Figure \ref{fig1} and Table \ref{table:model_config} respectively illustrate the architectures of the three upstream feature extractor paradigms and their corresponding fundamental configurations. The backend MOS predictor processes the C-dimensional hidden representations through a feed-forward network with a hidden size of $4096$, utilizing ReLU activation and $0.1$ dropout. To capture temporal dependencies, we employ 4 multi-head attention layers with 8 heads and a dropout rate of $0.2$.  We use the SGD \cite{Bottou2012SGD} optimizer with a fixed learning rate of $1 \times 10^{-4}$, momentum of $0.9$, and a batch size of $2$. The training spans $30$ epochs using L2 loss, incorporating an early stopping mechanism with a patience of $20$ epochs. The checkpoint with the highest SRCC on the validation set is selected. To ensure reliability, results are averaged over three runs with different random seeds.

\noindent We use Mean Squared Error (MSE), Linear Correlation Coefficient (LCC), Spearman Rank Correlation Coefficient (SRCC), and Kendall Rank Correlation Coefficient (KTAU) to evaluate MOS prediction performance.

\begin{figure}[t]
    \centering
    \includegraphics[width=0.96\linewidth]{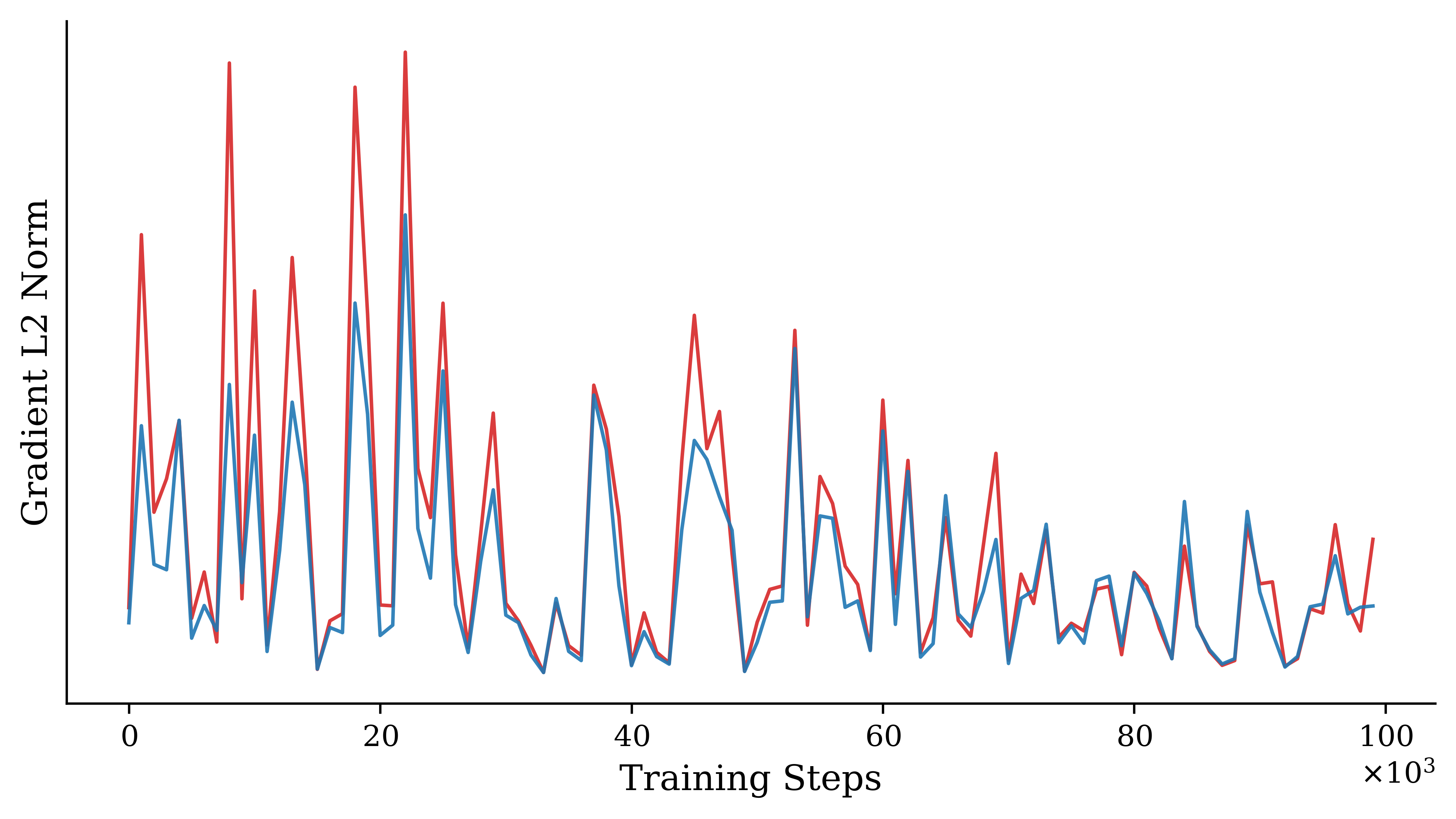}
    \vspace{-2mm} 
    \caption{Average L2 norm of gradients for semantic (red) and acoustic (blue) modules during fine-tuning, indicating the relative contribution of each module to training.}
    \label{fig2}
\end{figure}
\vspace{-1.5mm}
\begin{table*}[t]
  \centering
  \footnotesize 
  \renewcommand{\arraystretch}{0.95}
  \setlength{\tabcolsep}{3.2pt} 
  
  \definecolor{ssl_bg}{RGB}{235, 245, 255} 
  \definecolor{nac_bg}{RGB}{255, 240, 240} 
  \definecolor{uni_bg}{RGB}{240, 255, 240} 
  \definecolor{panel_bg}{gray}{0.9}

  \caption{Main results of MOS prediction on the BVCC dataset. The \textbf{bold} and \underline{underlined} numbers indicate the optimal and sub-optimal results, respectively.}
  \label{table:main results}
  
  \begin{tabular}{c l c c c c c c c c} 
   \toprule
   \multirow{2}{*}{\textbf{Type}} & \multirow{2}{*}{\textbf{Model}} & \multicolumn{4}{c}{\textbf{Frozen}} & \multicolumn{4}{c}{\textbf{Fine-Tuning}} \\
   \cmidrule(lr){3-6} \cmidrule(lr){7-10} 
    & & MSE$\downarrow$ & LCC$\uparrow$ & SRCC$\uparrow$ & KTAU$\uparrow$ & MSE$\downarrow$ & LCC$\uparrow$ & SRCC$\uparrow$ & KTAU$\uparrow$ \\
   \midrule
   
   \rowcolor{ssl_bg} & w2v2\_base & 0.261 & 0.839 & 0.836 & 0.652 & 0.267 & 0.879 & 0.877 & 0.703 \\
   \rowcolor{ssl_bg} & hubert\_base & 0.296 & 0.835 & 0.833 & 0.650 & 0.257 & \underline{0.880} & 0.878 & \underline{0.704} \\
   \rowcolor{ssl_bg} & hubert\_large & 0.304 & 0.809 & 0.810 & 0.626 & 0.226 & 0.875 & 0.874 & 0.699 \\ 
   \rowcolor{ssl_bg} \multirow{-4}{*}{\textbf{SSL}} & wavlm\_base & \underline{0.239} & \textbf{0.862} & \textbf{0.863} & \textbf{0.683} & \underline{0.210} & 0.877 & 0.875 & 0.700 \\

   \midrule
   \rowcolor{nac_bg} & EnCodec & 0.473 & 0.678 & 0.666 & 0.488 & 0.437 & 0.710 & 0.701 & 0.519 \\
   \rowcolor{nac_bg} \multirow{-2}{*}{\textbf{NAC}} & DAC & 0.405 & 0.726 & 0.715 & 0.531 & 0.391 & 0.748 & 0.740 & 0.555 \\
   
   \midrule
   \rowcolor{uni_bg} & Xcodec-hubert & 0.263 & 0.856 & 0.855 & 0.674 & 0.364 & 0.874 & \underline{0.879} & \underline{0.704} \\
   \rowcolor{uni_bg} & Xcodec-wavlm & \underline{0.239} & 0.859 & \underline{0.859} & \underline{0.680} & \textbf{0.201} & \textbf{0.882} & \textbf{0.882} & \textbf{0.710} \\
   \rowcolor{uni_bg} \multirow{-3}{*}{\textbf{Unified NAC}} & SpeechTokenizer & \textbf{0.233} & \underline{0.860} & 0.856 & 0.676 & 0.226 & 0.872 & 0.872 & 0.696 \\
   \bottomrule
  \end{tabular}
\end{table*}

\begin{table*}[t]
  \centering
  \footnotesize 
  \renewcommand{\arraystretch}{0.95}
  \setlength{\tabcolsep}{3.2pt} 
  
  \definecolor{ssl_bg}{RGB}{235, 245, 255} 
  \definecolor{nac_bg}{RGB}{255, 240, 240} 
  \definecolor{uni_bg}{RGB}{240, 255, 240} 
  \definecolor{panel_bg}{gray}{0.9}

  \caption{Zero-shot generalization results on OOD datasets. The \textbf{bold} and \underline{underlined} numbers indicate the optimal and sub-optimal results, respectively.}
  \label{table:ood_results}
  
  \begin{tabular}{c l c c c c c c c c} 
      \toprule
      \multirow{2}{*}{\textbf{Type}} & \multirow{2}{*}{\textbf{Model}} & \multicolumn{4}{c}{\textbf{Frozen}} & \multicolumn{4}{c}{\textbf{Fine-Tuning}} \\
      \cmidrule(lr){3-6} \cmidrule(lr){7-10} 
      & & MSE$\downarrow$ & LCC$\uparrow$ & SRCC$\uparrow$ & KTAU$\uparrow$ & MSE$\downarrow$ & LCC$\uparrow$ & SRCC$\uparrow$ & KTAU$\uparrow$ \\
      
      \midrule
      \multicolumn{10}{c}{\cellcolor{panel_bg}\textbf{SOMOS (cross-corpus)}} \\ 
      \midrule
      
      \rowcolor{ssl_bg} & w2v2\_base & 0.724 & 0.326 & 0.317 & 0.216 & 0.573 & 0.400 & 0.390 & 0.267 \\
      \rowcolor{ssl_bg} & hubert\_base & 1.013 & 0.318 & 0.310 & 0.211 & 0.570 & 0.350 & 0.344 & 0.234 \\
      \rowcolor{ssl_bg} \multirow{-3}{*}{\textbf{SSL}} & wavlm\_base & \underline{0.480} & \underline{0.419} & 0.404 & 0.277 & 0.626 & 0.352 & 0.360 & 0.249 \\

      \rowcolor{uni_bg} & Xcodec-hubert & 0.627 & \textbf{0.450} & \textbf{0.455} & \textbf{0.311} & 0.413 & 0.383 & 0.366 & 0.250 \\
      \rowcolor{uni_bg} & Xcodec-wavlm & 0.518 & 0.329 & 0.309 & 0.210 & \underline{0.353} & \underline{0.410} & \underline{0.402} & \underline{0.276} \\
      \rowcolor{uni_bg} \multirow{-3}{*}{\textbf{Unified NAC}} & SpeechTokenizer & \textbf{0.425} & \underline{0.419} & \underline{0.407} & \underline{0.279} & \textbf{0.323}& \textbf{0.444} & \textbf{0.440} & \textbf{0.302} \\

      \midrule
      \multicolumn{10}{c}{\cellcolor{panel_bg}\textbf{BC2019 (cross-lingual)}} \\ 
      \midrule
      
      \rowcolor{ssl_bg} & w2v2\_base & 2.872 & \underline{0.612} & \underline{0.644}& \underline{0.460} & \underline{2.314} & 0.627 & 0.653 & 0.471\\
      \rowcolor{ssl_bg} & hubert\_base & 3.233 & 0.538 & 0.576 & 0.420 & 2.811 & 0.590 & 0.618 & 0.444 \\
      \rowcolor{ssl_bg} \multirow{-3}{*}{\textbf{SSL}} & wavlm\_base & \textbf{1.859} & \textbf{0.647} & \textbf{0.674} & \textbf{0.478} & \textbf{1.966} & \underline{0.659} & \underline{0.671} & \textbf{0.493} \\
      
      \rowcolor{uni_bg} & Xcodec-hubert & 3.697 & 0.489 & 0.563 & 0.397 & 2.429 & 0.582 & 0.611 & 0.422 \\
      \rowcolor{uni_bg} & Xcodec-wavlm & \underline{2.597} & 0.599 & 0.617 & 0.440 & 2.360 & \textbf{0.665} & \textbf{0.682} & \underline{0.492} \\
      \rowcolor{uni_bg} \multirow{-3}{*}{\textbf{Unified NAC}} & SpeechTokenizer & 3.497 & 0.574 & 0.610 & 0.436 & 3.864 & 0.498 & 0.577 & 0.411 \\
      \bottomrule
    \end{tabular}
\end{table*}

\subsection{Experimental Results and Analysis}

Table \ref{table:main results} and Table \ref{table:ood_results} present the performance of different feature extractors under both frozen and fine-tuning protocols. Our analysis reveals three insights.

\noindent\textbf{Semantic-acoustic synergy is highly beneficial for intrinsic quality assessment.}
In the frozen setting on the BVCC dataset (Table \ref{table:main results}), models that synergize semantic understanding with fine-grained acoustic modeling (specifically unified NACs and WavLM) tend to establish a higher performance upper bound, outperforming standard semantic-focused SSLs (Wav2Vec 2.0 and HuBERT).
This performance stratification highlights potential limitations of purely semantic paradigms: the intrinsic representations of standard SSLs tend to discard paralinguistic details that can be crucial for precise MOS prediction.
In contrast, unified NACs address this by explicitly injecting semantic information into an acoustic backbone via fusion or distillation. Similarly, WavLM, while architecturally an SSL model, effectively integrates acoustic priors through its masked speech denoising objective. Both approaches support our hypothesis that acoustic-semantic synergy, rather than a singular focus on either, is highly beneficial for accurate quality perception. Conversely, the poor performance of acoustic-only NACs indicates that acoustic fidelity alone is often insufficient without high-level semantic guidance.

\noindent\textbf{The MOS prediction task is semantic-dominant.} In the fine-tuning setting, the performance gap between unified NACs and SSL models narrows considerably, indicating that the MOS prediction task is inherently semantic-dominant. Our gradient analysis corroborates this observation: across 100k training iterations (sampled at 1k-step intervals), the semantic module consistently yields a higher average gradient magnitude than the acoustic module, as visualized in Figure 2. Nevertheless, Xcodec-wavlm remains the top-performing model, which demonstrates that the joint semantic-acoustic representations deliver the highest performance upper bound. Notably, the superiority of SpeechTokenizer is slightly attenuated. We attribute this phenomenon to our implementation strategy: the discrete codebook and its corresponding SSL teacher model, which form the core semantic components of SpeechTokenizer, were not jointly optimized, thereby restricting its adaptive capacity.

\noindent\textbf{Generalization behaviors diverge across domain and language shifts.}
Table \ref{table:ood_results} highlights a distinct contrast in zero-shot generalization:
\begin{itemize}[leftmargin=*,labelsep=5pt]
\item \textbf{Cross-Corpus (Matched-language):} On the English-based SOMOS dataset, unified NACs consistently outperform SSL baselines across both frozen and fine-tuning configurations. This performance gap indicates that while SSL models excel at capturing linguistic naturalness, they often struggle with the distribution shifts of synthetic artifacts in specialized neural-synthesis samples. The robustness of unified NACs in this OOD scenario suggests the hypothesis that semantic representations alone may not be fully sufficient for universal quality assessment. In such cases, acoustic priors appear to provide critical sensitivity to fine-grained degradations that semantic-focused features might otherwise overlook.
\item \textbf{Cross-Lingual (Unseen-language):} On the Chinese-based BC2019 dataset, the advantage of unified NACs diminishes, with SSL models showing comparable or superior robustness. This performance shift originates from the representation space: SSLs utilize continuous latent spaces to capture supra-segmental features that transcend language boundaries, whereas unified NACs are constrained by discrete codebooks biased toward the pre-training language. When applied to a distinct language, phonotactic mismatches force the NAC encoder to project features into a suboptimal manifold, degrading quality. These results indicate that while acoustic-semantic integration excels in matched-language tasks, the high-level abstractions of SSLs remain more resilient to the distribution shifts in cross-lingual OOD scenarios.
\end{itemize}

\section{Conclusion}
This study systematically investigates the potential limitations of semantic representations for high-accuracy MOS prediction. Comparisons among SSLs, acoustic-only NACs, and unified NACs suggest that semantics alone may not fully capture speech quality, and incorporating fine-grained acoustic information appears to provide complementary benefits, improving the performance upper bound.
In our experiments, unified representations generally outperform single-modal models on the BVCC and SOMOS datasets. Although SSLs exhibit stronger cross-lingual generalization, multi-scale fusion appears beneficial for robust assessment.
Future work will explore ways to improve the cross-lingual adaptability of unified NACs to further guide the optimization of speech generation models..

\bibliographystyle{IEEEtran}

\bibliography{mybib}


\end{document}